\documentclass[aps,twocolumn,prl]{revtex4-1}
\usepackage[
pdfstartview={FitH},
bookmarks=true,
bookmarksnumbered=true,
bookmarksopen=true,
bookmarksopenlevel=\maxdimen,
pdfborder={0 0 0},
colorlinks=true,
linkcolor = blue,
citecolor=blue,
urlcolor=blue,
filecolor=black,
pdfauthor={Ferenc Huszar and Neil Houlsby},
pdftitle={Adaptive Bayesian Quantum Tomography},
pdfdisplaydoctitle=true
]{hyperref}
\usepackage{tikz}
\usepackage{pgfplots}
\usepackage{amsmath}
\usepackage{amssymb}
\usepackage{microtype}

\definecolor{mycolor1}{rgb}{1,0,1}

\newcommand{\param}{{\rho}} \newcommand{\data}{\mathcal{D}}
\newcommand{\argmax}{ \operatorname*{argmax}}
\newcommand{\config}{\alpha} \newcommand{\configset}{\mathcal{A}}
\newcommand{\outcome}{\gamma} 
\newcommand{\eg}{e.\,g.\ }

\begin{document}
\title{Adaptive Bayesian Quantum Tomography}

\author{F.~Husz\'{a}r}
\affiliation{Computational and Biological Learning Lab, Department of Engineering, University of Cambridge, Cambridge, UK}
\author{N.\,M.\,T.~Houlsby}
\affiliation{Computational and Biological Learning Lab, Department of Engineering, University of Cambridge, Cambridge, UK}

\begin{abstract}
In this letter we revisit the problem of optimal design of quantum tomographic experiments. In contrast to previous approaches where an optimal set of measurements is decided in advance of the experiment, we allow for measurements to be adaptively and efficiently re-optimised depending on data collected so far. We develop an adaptive statistical framework based on Bayesian inference and Shannon's information, and demonstrate a ten-fold reduction in the total number of measurements required as compared to non-adaptive methods, including mutually unbiased bases.
\end{abstract}

\maketitle

Quantum tomography is a valuable tool in quantum information processing, being essential for characterisation of quantum states, gates, and measurement equipment.  Quantum state tomography (QST) aims to determine an unknown quantum state from the outcome of measurements performed on an ensemble of identically prepared systems. Measurements in quantum systems are non-deterministic, hence QST is a classical statistical estimation problem. Full tomography is inherently resource-intensive: even in moderately sized systems, the number of measurements required is often prohibitive.
There is a need for methods that allow for shorter experiments. Optimal experiment design (OED) aims to achieve this by selecting cleverly which measurements to use during the experiment.

Most existing approaches to OED determine, prior to collecting data, an optimal set of measurements to be used throughout the experiment. In this sense, whenever they exist, mutually unbiased bases (MUBs) are known to be optimal\,\cite{MUBFirst,MUBExperiment}. Research since has focused mainly on proving or disproving existence of, and implement MUBs in various dimensions\,\cite{DimensionSix,MUBQutrit,MUBExperiment}. Other work,\,\cite{OEDFirst,OEDAverage} considered OED based on the Cram\'{e}r-Rao bound. Here we argue that these approaches, including MUBs, provide only a partial solution to the problem of optimal experiment design inasmuch as they do not take partial data into account. If we are allowed to revise our choice of measurements during the experiment based on data collected so far, we may be in a better position to reduce redundancy. This strategy is generally known as active learning or adaptive sampling. In physics, this approach has been referred to as self-learning measurements \cite{SelfLearning, SelfLearningExperimental}. However, due to the expensive computations that are involved, these methods have been restricted to two dimensional pure quantum states, or very few measurements. Recently advances in Bayesian methods allow us to build a fast, online algorithm that allows self-learning in arbitrary dimensions with many measurements.

Here we propose a new algorithmic framework that we call \emph{Adaptive Bayesian Quantum Tomography} (ABQT), that builds on full Bayesian inference and Shannon information. 
To achieve adaptivity in practice, we need a fast algorithm for performing Bayesian state reconstruction from partial data after each measurement. Current sampling methods such as in \cite{BayesianTomography} are inappropriate as their costs increase with the number of measurement configurations tried so far. As a solution, we present a sequential importance sampling scheme\,\cite{SMCBook}, that does not suffer from this. We then use the developed algorithm in conjunction with an information theoretic objective to adaptively optimise measurements. We assess the relative performance of our adaptive method in Monte Carlo simulations of qubit systems, and demonstrate a ten-fold reduction in the number of measurements needed for full tomography of two-qubit pure states. We also investigate the trade off between entangling and separable measurements in multipartite systems. Our central finding is that via adaptive tomography one can achieve, and even surpass, the statistical efficiency of MUB tomography using only separable measurements, that require experimental apparatus that is substantially easier to build using current technology.

\paragraph{Quantum state tomography} involves determining from experimental data the quantum state, $\rho$, of a system by performing measurements on several identical copies. For a $D$-dimensional system ($D=2^m$ for $m$-qubit systems), $\rho$ is an $D \times D$ complex-valued density matrix. $\rho$ has to be Hermitian and have unit trace, so $D^2-1$ real degrees of freedom must be estimated. The apparatus for a tomographic experiment may be configured in several different ways; we use $\config\in\configset$ to index all accessible configurations. Each measurement configuration $\config$ is characterised by a positive operator-valued measure (POVM). For each configuration, a measurement results in observing one of a finite number, $\Gamma$, of distinguishable outcomes. A POVM is defined by a set, $\mathbb{M}_{\config}$, of Hermitian operators $M_{\config\outcome}$, indexed by possible outcomes $\outcome\in\{1,\ldots,\Gamma\}$, satisfying $\sum_{\outcome=1}^{\Gamma} M_{\config\outcome} = I$. These POVMs jointly constitute our tomographic model $\mathcal{M}=\{\mathbb{M}_{\config}:\config\in\configset\}$ and determine the probability of observing outcome $\outcome$ in configuration $\config$ when the measured system is in state $\param$ via Born's rule:

\begin{equation}
\mathbb{P}\left(\gamma\vert\param,\config;\mathcal{M}\right) = \text{tr}\left\{M_{\config\outcome}\param \right\}\label{eqn:born}\notag
\end{equation}

State reconstruction has been approached with several methods, the most popular being maximum likelihood estimation (MLE). MLE finds a physically feasible state $\param$ that is most likely to have produced the observed data, $\mathcal{D}$, by maximising the likelihood:

\begin{equation}
\label{eqn:lik}
\mathcal{L}(\rho;\data)=\prod_{n=1}^{N} \mathbb{P}\left(\outcome_n\vert\param,\config_n\right) = \prod_{\config\in\configset}\prod_{\gamma=1}^{\Gamma}\mbox{tr}\{M_{\alpha\gamma}\rho\}^{c_{\alpha\gamma}}
\end{equation}

where $c_{\alpha\gamma}$ is the number of times outcome $\outcome$ was observed in configuration $\config$. 
All probabilities are conditional on $\mathcal{M}$, for brevity this is omitted. A well-known drawback of MLE is that it often yields rank-deficient estimates, and thus assigns zero predictive probability to certain observations\,\cite{BayesianTomography}. This seems an unreasonable conclusion on the basis of a finite sample.
Additionally, MLE provides no measure of uncertainty in its point estimate.

More sophisticated methods for quantum tomography use Bayesian inference and suffer from neither of these problems\,\cite[][and refs.]{BayesianTomography}. In Bayesian inference a prior probability density, $p(\param)$, over feasible states is specified. This prior is then augmented with the likelihood from Eqn.\,\eqref{eqn:lik} using Bayes' rule to yield a posterior distribution:

\begin{equation}
\label{eqn:bayes}
p(\param\vert\data)\propto \mathcal{L}(\param;\data)p(\param)
\end{equation}

Should we want a point estimate, we may report, say, the Bayesian mean estimate (BME) which is known to maximise expected operational divergences\,\cite{BayesianTomography,BayesianOptimality}. But importantly, Bayesian inference also provides \emph{error bars}, and more: the posterior captures richly our remaining uncertainty in the true state having seen the data $\data$. 

For Bayesian inference one has to provide the prior $p(\param)$, which is typically chosen to be non-informative or uniform. Here we adopt the representation and prior introduced in\,\cite{BayesianTomography}, that treats our original system of interest as part of a larger, $D\times K$ dimensional bipartite system. Our prior over the mixed state $\param$ is then defined as the measure induced by the uniform (Haar) measure over pure states in $D\times K$ dimensions. It is easy to see that, tracing out the $K$ dimensional ancillary part leaves us with a rank-$K$ mixed state $\rho$. Thus, by tuning this parameter we can trade off between computational efficiency and estimation accuracy, in a similar manner to compressed sensing\,\cite{CompressedSensing}.

Unfortunately, normalisation of the posterior distribution (Eqn.\,\eqref{eqn:bayes}) becomes analytically intractable, and therefore we have to approximate it, usually via Markov chain Monte Carlo (MCMC) methods. Several MCMC approaches have been suggested in this context\,\cite[][and refs.\ therein]{BayesianTomography}. These methods require evaulation of the full likelihood \eqref{eqn:lik}, which has $\mathcal{O}(n)$ cost with the number of different configurations used so far. This is undesirable for adaptive tomography, where inference has to be performed after each measurement. To address this problem we developed a fast sequential importance sampling (SIS) algorithm, with $\mathcal{O}(1)$ likelihood evaluation cost. As we are not aware of this approach being used in the context of QST, we briefly explain the basic version below. The interested reader is referred to\,\cite{SMCBook} for a thorough overview.

In SIS, one keeps track of a number of samples, often called particles, $\param_s,\, (s=1\ldots S)$ and corresponding weights $w_s, \, \left( \sum_s w_s = 1 \right)$  which are updated sequentially, every time a new measurement is made. Assume that after $n$ measurements, having observed data $\data_n$, the particles and weights  $w^{(n)}_s$ constitute an approximation to the posterior:

\begin{equation}
	p(\param\vert\data_n)\approx \sum_{s=1}^{S}w^{(n)}_s\delta(\param-\param_s)\label{eqn:SISapprox}
\end{equation}

Using this approximation, and Bayes' rule, one can derive an approximation to the next posterior, after observing a new outcome $\outcome_{n+1}$ in configuration $\config_{n+1}$, as:

\begin{align}
	p(\param\vert\config_{n+1}&,\outcome_{n+1},\data_n) = \frac{\mathbb{P}(\outcome_{n+1}\vert\param,\config_{n+1})p(\param\vert\data_n)}{\int \mathbb{P}(\outcome_{n+1}\vert\param,\config_{n+1})p(\param\vert\data_n) d\param}\label{eqn:SISupdate}\\
	&\approx \sum_{s=1}^{S}\underbrace{\frac{\mathbb{P}(\outcome_{n+1}\vert\param_s,\config_{n+1})w^{(n)}_s}{\sum_{r=1}^{S}\mathbb{P}(\outcome_{n+1}\vert\param_r,\config_{n+1})w^{(n)}_r}}_{w^{(n+1)}_s}\delta(\param-\param_s)\notag
\end{align}

The new weights $w^{(n+1)}_s$ are the renormalised product of our current weights $w^{(n)}_s$ and observation probabilities $\mathbb{P}(\outcome_{n+1}\vert\param_s,\config_{n+1})$. This update is fast, and only requires computing one term of the full likelihood, thus its complexity is independent of how many configurations have been tried before. This computational efficiency comes at a price; as time progresses, several weights decay to almost zero, and thus the quality of our approximation drops. This issue can be detected and handled by monitoring the effective sample size and resampling appropriately\,\cite{SMCBook}.

Having discussed our method for estimating the state based on partial data, we now turn to the problem of optimal experiment design. Different state determination schemes have different OED strategies associated with them. Maximum likelihood methods usually use some form of the Cram\'{e}r-Rao bound\,\cite{OEDFirst,OEDAverage}. Bayesian experiment design on the other hand is based on Shannon information\,\cite{MUBFirst,ExactInformation}. The posterior characterises our remaining uncertainty in the parameter, and this uncertainty can be quantified using Shannon's entropy. A sensible aim is to pick an experimental configuration $\config$, such that after observing the outcome $\outcome$, the entropy $\mathbb{H}$ of the new posterior is reduced the most:

\begin{equation}
\argmax_{\config\in\configset} \left\{ \mathbb{H}\left[p(\param|\data)\right] - \mathbb{E}_{p(\outcome\vert\config,\data)}\left[\mathbb{H}\left[ p(\param|\outcome,\config,\data) \right] \right]\right\}\label{eqn:expectedentropyreduction}
\end{equation}

The expectation with respect to $\outcome$ is needed as the measurement outcome is unknown \emph{a priori}. This objective naturally allows us to address the question `Having seen the outcome of the first few measurements, which measurement should we carry out next?'
Rather, it was used to determine a single best set of measurements which are then uniformly sampled throughout the experiment\,\cite{MUBFirst,ExactInformation}. Under these circumstances mutually unbiased bases (MUBs) are optimal, whenever they exist. We exploit the dependence of Eqn.\,\eqref{eqn:expectedentropyreduction} on past observations, and allow for measurements to be re-optimised adaptively as the experiment progresses.

However, Eqn.\,\eqref{eqn:expectedentropyreduction}  is impractical to work with directly, as it involves computing entropies of high-dimensional intractable posterior densities. Recall that we approximate our posterior by samples, with which it is notoriously hard to estimate differential entropies. Furthermore, in Eqn. \eqref{eqn:expectedentropyreduction} the posterior has to be re-computed for every possible outcome $\outcome$. Therefore, instead of working with Eqn.\,\eqref{eqn:expectedentropyreduction} directly, we propose to use an equivalent reformulation thereof in terms of predictive distributions\,\cite{ExactInformation}:

\begin{equation}
\argmax_{\config\in\configset} \left\{ \mathbb{H}\left[\mathbb{P}(\outcome|\config,\data)\right] - \mathbb{E}_{p(\param\vert\data)}\left[\mathbb{H}\left[ \mathbb{P}(\outcome|\config,\param) \right] \right]\right\}\label{eqn:rearrangement}
\end{equation}

In previous studies \cite{SelfLearning} the system is limited to pure single qubit states, calculating the intractable Bayesian normalising constant can be realised with simple numerical integration; this could not be extended easily to higher dimensions. They consider two active learning algorithms: firstly, uncertainty sampling, which uses an approximate version of Eqn. \eqref{eqn:rearrangement}, where the second term was ignored. This arguably leads to suboptimal selection behaviour; the experimenter's uncertainty may be confounded with inherent uncertainty of quantum measurements. The second seeks to minimize the Bayes Risk, using fidelity as the loss function; this requires a posterior update for evaluation of every measurement to be considered, ABQT requires only one update per complete cycle. Online computation is therefore infeasible, in \cite{SelfLearningExperimental} experimental designs for all $2^N$ possible experimental outcome successions are pre-computed, they are therefore limited to very short experiments ($< 20$ measurements). Combining Eqn. \eqref{eqn:rearrangement} with our SIS Bayesian update scheme allows for fast online experimental design.

\begin{figure}
\resizebox{.9\columnwidth}{!}{
\includegraphics{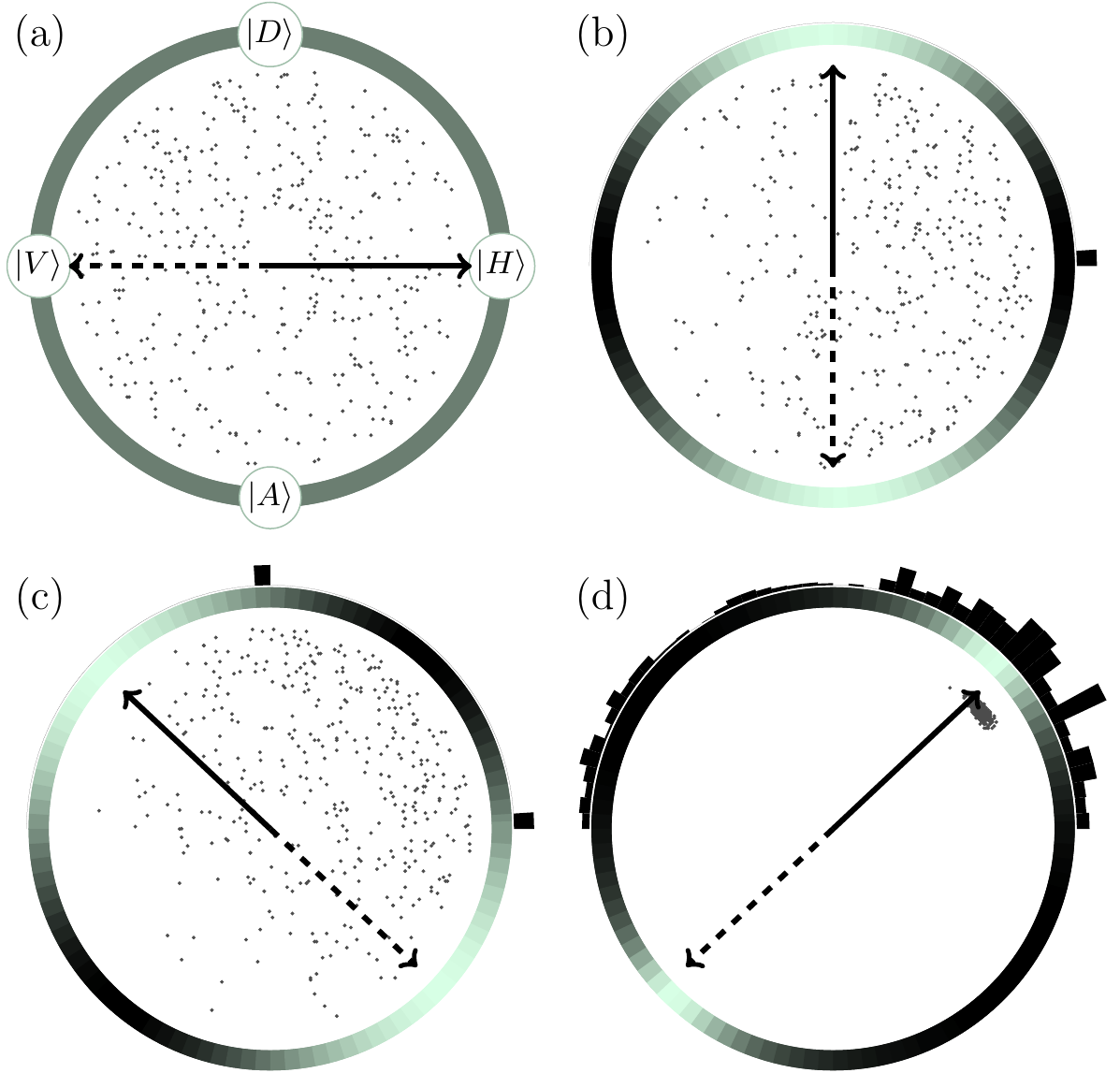}
}

\caption{Adaptive selection of measurements based of partial data. Scatter plots show 400 samples from current posterior. Shaded circles around the `Bloch disk' show relative value of the objective in Eqn.\,\eqref{eqn:rearrangement} for different measurement directions (lighter is higher). Pairs of arrows show the most informative next measurement. Circular histograms show the number of times measurement directions have been used. \textbf{(a)}  Initially, no observations are made, samples shown are from the uniform prior. All measurements are equally informative, we chose to start with $\{\left\vert H\right\rangle,\left\vert V\right\rangle\}$. \textbf{(b)}  After one measurement, the posterior is updated, the next best measurement is mutually unbiased w.r.t.\ the first one. It is now $\{\left\vert D\right\rangle,\left\vert A\right\rangle\}$. \textbf{(c)} After two observations, the next best measurement is equally biased to the first two bases. \textbf{(d)} Posterior after 1000 observations concentrates around true state. The method tries a range of measurements, with a tendency to point towards the solution.
\label{fig:Bloch disk}}
\end{figure}

\begin{figure}
%
%
\begin{tikzpicture}

\begin{loglogaxis}[%
scale only axis,
width=1.2in,
height=1in,
xmin=100, xmax=10000,
ymin=0.001, ymax=0.1,
xlabel={no.\ measurements ($n$)},
ylabel={mean infidelity},
axis on top
,anchor=north west]
\addplot [
color=black,
dotted,
line width=1.0pt
]
coordinates{ (100,0.0936504) (127,0.0831784) (162,0.0703672) (206,0.0552666) (263,0.0442545) (335,0.0367459) (428,0.0288689) (545,0.0240754) (695,0.0219549) (885,0.0182243) (1128,0.0154802) (1438,0.0127685) (1832,0.00966868) (2335,0.00797919) (2976,0.00641328) (3792,0.00604222) (4832,0.00543209) (6158,0.0051745) (7847,0.0052069) (10000,0.00427943)
};
\label{leg:rand}

\addplot [
color=red,
dashed,
line width=1.0pt
]
coordinates{ (100,0.0854107) (127,0.0685189) (162,0.0556463) (206,0.0536513) (263,0.0455832) (335,0.0361215) (428,0.0293875) (545,0.0261447) (695,0.0222585) (885,0.0205318) (1128,0.0170572) (1438,0.0149583) (1832,0.0132948) (2335,0.0108542) (2976,0.00907246) (3792,0.00787257) (4832,0.0070859) (6158,0.00582692) (7847,0.00538335) (10000,0.00451835)
};

\addplot [
color=blue,
dash pattern=on 4pt off 2pt on 1pt off 2pt,
line width=1.0pt
]
coordinates{ (100,0.0770638) (127,0.066016) (162,0.0639258) (206,0.0575981) (263,0.0488788) (335,0.0416774) (428,0.0385588) (545,0.0306082) (695,0.0264267) (885,0.0209842) (1128,0.0168218) (1438,0.0157689) (1832,0.0132344) (2335,0.0124372) (2976,0.0110141) (3792,0.0101086) (4832,0.00830792) (6158,0.00739129) (7847,0.00680199) (10000,0.00627426)
};

\addplot [
color=green,
solid,
line width=1.5pt
]
coordinates{ (100,0.0717411) (127,0.0634239) (162,0.0479571) (206,0.0386032) (263,0.0300618) (335,0.0213717) (428,0.0172278) (545,0.0121796) (695,0.0100679) (885,0.00768467) (1128,0.0064199) (1438,0.00571155) (1832,0.00436206) (2335,0.00366658) (2976,0.0030864) (3792,0.0025154) (4832,0.00203493) (6158,0.00194272) (7847,0.00142223) (10000,0.00119099)
};
\label{leg:aFlex}

\end{loglogaxis}

%
%

\begin{semilogyaxis}[%
at={(1.65in,0)},
scale only axis,
width=1.1in,
height=1in,
xmin=0.5, xmax=1,
xtick={0.5,0.75,1},
ymin=0.0005, ymax=0.01,
xlabel={purity},
axis on top
,anchor=north west]
\addplot [
color=black,
dotted,
line width=1.0pt
]
coordinates{ (0.5,0.00075929) (0.6,0.000857054) (0.7,0.000956602) (0.8,0.000893443) (0.85,0.00111663) (0.9,0.00113761) (0.925,0.00221566) (0.95,0.0024295) (0.975,0.00216895) (1,0.00581406)
};

\addplot [
color=red,
dashed,
line width=1.0pt
]
coordinates{ (0.5,0.000728511) (0.6,0.000663376) (0.7,0.000813218) (0.8,0.000983281) (0.85,0.00146818) (0.9,0.00145583) (0.925,0.00191827) (0.95,0.00189825) (0.975,0.00225826) (1,0.00709776)
};

\addplot [
color=blue,
dash pattern=on 4pt off 2pt on 1pt off 2pt,
line width=1.0pt
]
coordinates{ (0.5,0.000933037) (0.6,0.000803118) (0.7,0.00094978) (0.8,0.000968539) (0.85,0.00136559) (0.9,0.00113676) (0.925,0.00114692) (0.95,0.0016187) (0.975,0.00198486) (1,0.00877819)
};

\addplot [
color=green,
solid,
line width=1.5pt
]
coordinates{ (0.5,0.000583439) (0.6,0.000722364) (0.7,0.000871828) (0.8,0.000775065) (0.85,0.000892919) (0.9,0.000805987) (0.925,0.00087682) (0.95,0.00094804) (0.975,0.00114797) (1,0.00169893)
};

\end{semilogyaxis}

\node at (1.05in,-0.13in) {(a)};
\node at (1.80in,-0.13in) {(b)};

\end{tikzpicture}
\caption{One qubit tomography using projective measurements. \textbf{(a)}  Improvement of mean posterior fidelity as the experiment progresses. Results are shown for uniformly sampled measurements (\ref{leg:rand}), uniformly sampled Pauli measurements ( \ref{leg:MUB}), ABQT selecting adaptively amongst the 3 Pauli measurements (\ref{leg:aMUB}) and ABQT picking general measurements (\ref{leg:aFlex}). Adaptive optimisation of measurements allows for an almost $n^{-1}$ rate of convergence, while other methods are more consistent with a $n^{-\frac{1}{2}}$ rate. \textbf{(b)}  Final value of the mean posterior infidelity after 6000 measurements using the four methods as before, as a function of purity of the state to be estimated. The advantage of ABQT is greatest for purer states. \label{fig:qubit_results}}
\end{figure}
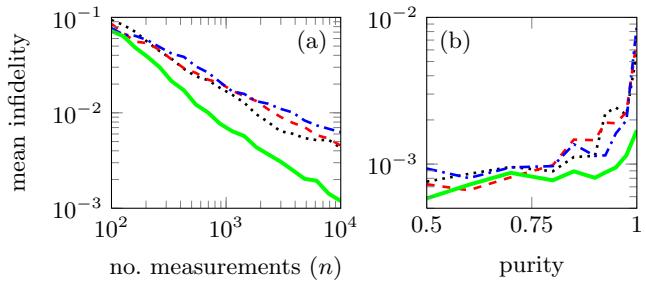

\begin{figure*}[th]
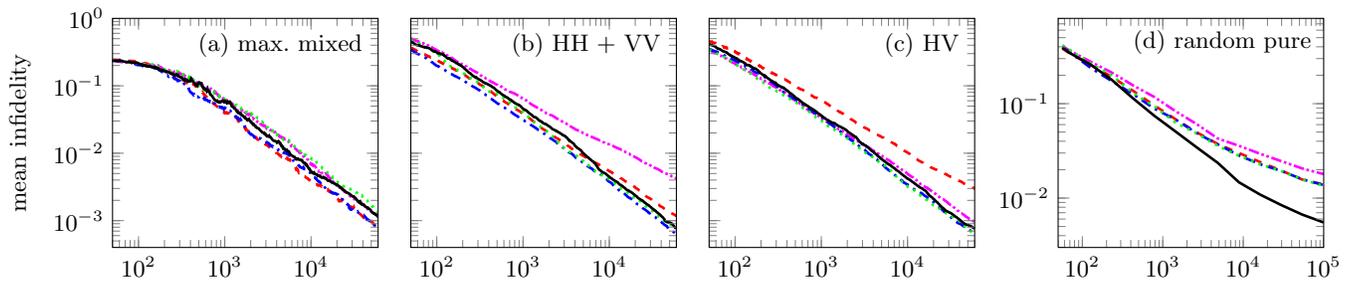

\input{figures/case1.tikz}
\input{figures/case2.tikz}
\input{figures/case3.tikz}
%
%
\begin{tikzpicture}

\definecolor{mycolor1}{rgb}{1,0,1}

\begin{loglogaxis}[%
scale only axis,
width= 1.39in,
height= 1.2in,
xmin=50, xmax=100000,
ymin=0.003, ymax=0.8,
axis on top]
\addplot [
color=red,
dashed,
line width=1.0pt
]
coordinates{ (56.5,0.383409) (69,0.342484) (127,0.256494) (233,0.192413) (428,0.136535) (784,0.0958473) (1438,0.0685589) (2636,0.0497711) (4832,0.0373716) (8858,0.0300054) (16237,0.0240709) (29763,0.0195324) (54555,0.0156758) (100000,0.0139577)
};

\addplot [
color=mycolor1,
dash pattern=on 4pt off 1pt on 1pt off 1pt on 1pt off 1pt,
line width=1.0pt
]
coordinates{ (56.5,0.407011) (69,0.365888) (127,0.277832) (233,0.212963) (428,0.156079) (784,0.117747) (1438,0.0832641) (2636,0.0607293) (4832,0.0427458) (8858,0.0357814) (16237,0.0294461) (29763,0.02459) (54555,0.0205416) (100000,0.017947)
};

\addplot [
color=blue,
dash pattern=on 4pt off 2pt on 1pt off 2pt,
line width=1.0pt
]
coordinates{ (56.5,0.392038) (69,0.352813) (127,0.239161) (233,0.173074) (428,0.126165) (784,0.0885277) (1438,0.068068) (2636,0.0488063) (4832,0.0370792) (8858,0.0282738) (16237,0.0228435) (29763,0.0193612) (54555,0.0157973) (100000,0.0137361)
};

\addplot [
color=green,
dotted,
line width=1.0pt
]
coordinates{ (56.5,0.416317) (69,0.372928) (127,0.267541) (233,0.199025) (428,0.129936) (784,0.0937346) (1438,0.0631296) (2636,0.0484169) (4832,0.0357245) (8858,0.0280432) (16237,0.0228003) (29763,0.0193429) (54555,0.0156744) (100000,0.0135143)
};

\addplot [
color=black,
solid,
line width=1.0pt
]
coordinates{ (56.5,0.39154) (69,0.345328) (127,0.259211) (233,0.178253) (428,0.113571) (784,0.0744741) (1438,0.0507662) (2636,0.0344829) (4832,0.0235882) (8858,0.0146908) (16237,0.0108828) (29763,0.00842673) (54555,0.00666891) (100000,0.00552)
};

\node[anchor=east] at (axis cs:90000,0.45) {(d) random pure};

\end{loglogaxis}
\end{tikzpicture}\\
\caption{Two qubit QST with uniformly chosen amongst MUB (\ref{leg:MUB}) or SSQT bases (\ref{leg:SSQT}) and ABQT picking from the same set of MUBs (\ref{leg:aMUB}),  SSQT bases (\ref{leg:aSSQT}) or a more flexible set of 81 separable bases (\ref{leg:fSSQT}). Cases (a)-(c) are the same as those in\,\cite{MUBExperiment}, (d) shows average results over 20 randomly generated entangled pure states. \textbf{(a)} As expected, for the maximally mixed state the choice of measurement strategy has little effect. \textbf{(b)} On the entangled state $(\vert HH\rangle+\vert VV\rangle)/\sqrt{2}$ MUB outperforms SSQT when uniformly sampled, but by allowing for adaptivity we can close the performance gap. \textbf{(c)} SSQT outperforms MUBs on the separable state $\vert HV \rangle$, but again, picking measurements adaptively the two sets perform similarly. \textbf{(d)} For random pure states a large improvement in performance is made when performing ABQT with the flexible set of separable measurements. Using this set, ABQT only needs $10^4$ measurements to achieve $\approx98.7\%$ mean fidelity for which MUB needs $10^5$ measurements. \label{fig:two_qubit_results}}
\end{figure*}

The equivalence between Eqns.\, \eqref{eqn:expectedentropyreduction} and \eqref{eqn:rearrangement} becomes clear realising that they both express the conditional mutual information between $\param$ and $\outcome$. Eqn.\,\eqref{eqn:rearrangement} offers computational advantages over Eqn.\,\eqref{eqn:expectedentropyreduction}: it only involves computing discrete entropies $\mathbb{H}\left[ \mathbb{P}(\outcome|\config,\param) \right]$ and expectations of these under the posterior. This objective function is generally non-convex in $\config$, but its value - and derivatives with respect to $\config$ - can now be efficiently computed using our weighted posterior samples from Eqn.\,\eqref{eqn:SISapprox}, allowing us to find the most informative $\config$ by direct optimisation.

In summary, we propose the following algorithm, called Adaptive Bayesian Quantum Tomography. After each single measurement, ABQT updates its approximate posterior using Eqn.\,\eqref{eqn:SISupdate}, then chooses the next measurement by direct numerical maximisation of the information theoretic objective in Eqn.\,\eqref{eqn:rearrangement}.

\paragraph{EX 1: single qubit tomography.} In our first simulated experiments we study tomography of single qubits ($D=2$). Mixed state qubits have three real degrees of freedom, $\rho$ is represented as a point in a unit ball, called the Bloch sphere. For illustration purposes we first omit the third component, and only infer two remaining parameters, which will lie in a unit (Bloch) disk. This corresponds to \eg determining linear polarisation of a photon, assuming that the circular polarisation is zero. We allow for arbitrary projective measurements with binary ($\Gamma = 2$) outcomes. These are represented by pairs of antipodal points on the perimeter of the Bloch disk. Now $\config\in[0,\pi)$ codes for the orientation. Fig.\ \ref{fig:Bloch disk} shows the progression of measurement bases chosen by ABQT. The first two measurements are mutually unbiased, however, the third measurement is equally biased with respect to both previous bases, demonstrating that using a fixed MUB set is suboptimal in the adaptive framework. Throughout the rest of the experiment the algorithm explores a wide range of measurements.

Fig.\ \ref{fig:qubit_results} shows that the posterior mean fidelity - this time inferring all three coordinates in the full Bloch sphere - improves at a faster rate when measurements are adaptively optimised. We quantify performance as mean posterior fidelity, rather than the fidelity of the Bayesian mean estimate, as the latter gives no indication of the confidence in our estimate. The rate is more consistent with a $n^{-1}$ law rather than $n^{-\frac{1}{2}}$ as predicted for non-adaptive methods\,\cite[][and refs.]{MUBExperiment}. Fig.\ \ref{fig:qubit_results}.b shows a larger advantage for states of high purity (defined as sum of squared eigenvalues).

\paragraph{EX 2: Separable vs.\ MUB tomography of two qubits.} In multipartite systems, such as $m$-qubit registers, there are two fundamentally different classes of measurements one can apply: separable or entangling. Separable tomographic experiments are straightforward and cheap to implement, while entangling measurements are statistically more powerful. Notably, entanglement is required for implementing MUBs. These differences are discussed extensively in\,\cite{MUBExperiment}. To investigate this trade-off in the light of adaptive tomography, we reproduce and extend the experiments in\,\cite{MUBExperiment}. Results are shown in Fig.\ \ref{fig:two_qubit_results}. Notably, all substantial differences between MUB and standard separable tomography (SSQT) vanish as we allow for adaptivity (Fig.\ \ref{fig:two_qubit_results}.a--c). Furthermore, for random pure states, we are able to realise a ten-fold improvement over MUBs when using flexible separable measurements (Fig.\ \ref{fig:two_qubit_results}.d). The results indicate that allowing for adaptivity with an imperfect, but flexible set of measurements offers greater advantages than using a fixed set of MUBs.


In summary, we have presented a new adaptive optimal experimental design framework and method based on Bayesian inference and Shannon's information. We showed that mutually unbiased bases, widely accepted as \emph{the} optimal measurements, represent only a partial solution and are suboptimal in the adaptive framework. Moreover, the adaptive framework applies regardless of dimensionality, and can be applied to spaces where MUBs do not even exist\,\cite{DimensionSix,ExactInformation}. This motivates a shift in experimental focus from implementing complex entangling measurements to implementing quickly reconfigurable simpler measurements. In quantum optics, this could be feasibly achieved via mechanically or electronically controlled liquid crystal wave plates.

\begin{acknowledgments}
We thank our advisors M Lengyel, Z Ghahramani and CE Rasmussen as well as G Cs\'{a}nyi, S Lacoste-Julien, E Snelson and S Strelchuk. We are supported by EPSRC and Trinity College\,(FH) and Google Europe\,(NMTH).
\end{acknowledgments}

\bibliography{quantum_bald}

\begin{thebibliography}{13}%
\makeatletter
\providecommand \@ifxundefined [1]{%
 \@ifx{#1\undefined}
}%
\providecommand \@ifnum [1]{%
 \ifnum #1\expandafter \@firstoftwo
 \else \expandafter \@secondoftwo
 \fi
}%
\providecommand \@ifx [1]{%
 \ifx #1\expandafter \@firstoftwo
 \else \expandafter \@secondoftwo
 \fi
}%
\providecommand \natexlab [1]{#1}%
\providecommand \enquote  [1]{``#1''}%
\providecommand \bibnamefont  [1]{#1}%
\providecommand \bibfnamefont [1]{#1}%
\providecommand \citenamefont [1]{#1}%
\providecommand \href@noop [0]{\@secondoftwo}%
\providecommand \href [0]{\begingroup \@sanitize@url \@href}%
\providecommand \@href[1]{\@@startlink{#1}\@@href}%
\providecommand \@@href[1]{\endgroup#1\@@endlink}%
\providecommand \@sanitize@url [0]{\catcode `\\12\catcode `\$12\catcode
  `\&12\catcode `\#12\catcode `\^12\catcode `\_12\catcode `\%12\relax}%
\providecommand \@@startlink[1]{}%
\providecommand \@@endlink[0]{}%
\providecommand \url  [0]{\begingroup\@sanitize@url \@url }%
\providecommand \@url [1]{\endgroup\@href {#1}{\urlprefix }}%
\providecommand \urlprefix  [0]{URL }%
\providecommand \Eprint [0]{\href }%
\providecommand \doibase [0]{http://dx.doi.org/}%
\providecommand \selectlanguage [0]{\@gobble}%
\providecommand \bibinfo  [0]{\@secondoftwo}%
\providecommand \bibfield  [0]{\@secondoftwo}%
\providecommand \translation [1]{[#1]}%
\providecommand \BibitemOpen [0]{}%
\providecommand \bibitemStop [0]{}%
\providecommand \bibitemNoStop [0]{.\EOS\space}%
\providecommand \EOS [0]{\spacefactor3000\relax}%
\providecommand \BibitemShut  [1]{\csname bibitem#1\endcsname}%
\let\auto@bib@innerbib\@empty
\bibitem [{\citenamefont {Wootters}\ and\ \citenamefont
  {Fields}(1989)}]{MUBFirst}%
  \BibitemOpen
  \bibfield  {author} {\bibinfo {author} {\bibfnamefont {W.}~\bibnamefont
  {Wootters}}\ and\ \bibinfo {author} {\bibfnamefont {B.}~\bibnamefont
  {Fields}},\ }\href {\doibase DOI: 10.1016/0003-4916(89)90322-9} {\bibfield
  {journal} {\bibinfo  {journal} {Ann.\ Phys.}\ }\textbf {\bibinfo {volume}
  {191}},\ \bibinfo {pages} {363 } (\bibinfo {year} {1989})}\BibitemShut
  {NoStop}%
\bibitem [{\citenamefont {Adamson}\ and\ \citenamefont
  {Steinberg}(2010)}]{MUBExperiment}%
  \BibitemOpen
  \bibfield  {author} {\bibinfo {author} {\bibfnamefont {R.~B.~A.}\
  \bibnamefont {Adamson}}\ and\ \bibinfo {author} {\bibfnamefont {A.~M.}\
  \bibnamefont {Steinberg}},\ }\href {\doibase 10.1103/PhysRevLett.105.030406}
  {\bibfield  {journal} {\bibinfo  {journal} {Phys.\ Rev.\ Lett.}\ }\textbf
  {\bibinfo {volume} {105}},\ \bibinfo {pages} {030406} (\bibinfo {year}
  {2010})}\BibitemShut {NoStop}%
\bibitem [{\citenamefont {Raynal}\ \emph {et~al.}(2011)\citenamefont {Raynal},
  \citenamefont {L\"u},\ and\ \citenamefont {Englert}}]{DimensionSix}%
  \BibitemOpen
  \bibfield  {author} {\bibinfo {author} {\bibfnamefont {P.}~\bibnamefont
  {Raynal}}, \bibinfo {author} {\bibfnamefont {X.}~\bibnamefont {L\"u}}, \ and\
  \bibinfo {author} {\bibfnamefont {B.-G.}\ \bibnamefont {Englert}},\ }\href
  {\doibase 10.1103/PhysRevA.83.062303} {\bibfield  {journal} {\bibinfo
  {journal} {Phys.\ Rev.\ A}\ }\textbf {\bibinfo {volume} {83}},\ \bibinfo
  {pages} {062303} (\bibinfo {year} {2011})}\BibitemShut {NoStop}%
\bibitem [{\citenamefont {Yan}\ \emph {et~al.}(2010)\citenamefont {Yan},
  \citenamefont {Yang},\ and\ \citenamefont {Cao}}]{MUBQutrit}%
  \BibitemOpen
  \bibfield  {author} {\bibinfo {author} {\bibfnamefont {F.}~\bibnamefont
  {Yan}}, \bibinfo {author} {\bibfnamefont {M.}~\bibnamefont {Yang}}, \ and\
  \bibinfo {author} {\bibfnamefont {Z.-L.}\ \bibnamefont {Cao}},\ }\href
  {\doibase 10.1103/PhysRevA.82.044102} {\bibfield  {journal} {\bibinfo
  {journal} {Phys. Rev. A}\ }\textbf {\bibinfo {volume} {82}},\ \bibinfo
  {pages} {044102} (\bibinfo {year} {2010})}\BibitemShut {NoStop}%
\bibitem [{\citenamefont {Kosut}\ \emph {et~al.}(2004)\citenamefont {Kosut},
  \citenamefont {Walmsley},\ and\ \citenamefont {Rabitz}}]{OEDFirst}%
  \BibitemOpen
  \bibfield  {author} {\bibinfo {author} {\bibfnamefont {R.}~\bibnamefont
  {Kosut}}, \bibinfo {author} {\bibfnamefont {I.~A.}\ \bibnamefont {Walmsley}},
  \ and\ \bibinfo {author} {\bibfnamefont {H.}~\bibnamefont {Rabitz}},\
  }\href@noop {} {\  (\bibinfo {year} {2004})},\ \Eprint
  {http://arxiv.org/abs/quant-ph/0411093} {arXiv:quant-ph/0411093} \BibitemShut
  {NoStop}%
\bibitem [{\citenamefont {{Nunn \emph{et al}}}(2010)}]{OEDAverage}%
  \BibitemOpen
  \bibfield  {author} {\bibinfo {author} {\bibfnamefont {J.}~\bibnamefont
  {{Nunn \emph{et al}}}},\ }\href {\doibase 10.1103/PhysRevA.81.042109}
  {\bibfield  {journal} {\bibinfo  {journal} {Phys.\ Rev.\ A}\ }\textbf
  {\bibinfo {volume} {81}},\ \bibinfo {pages} {042109} (\bibinfo {year}
  {2010})}\BibitemShut {NoStop}%
\bibitem [{\citenamefont {Fischer}\ \emph {et~al.}(2000)\citenamefont
  {Fischer}, \citenamefont {Kienle},\ and\ \citenamefont
  {Freyberger}}]{SelfLearning}%
  \BibitemOpen
  \bibfield  {author} {\bibinfo {author} {\bibfnamefont {D.~G.}\ \bibnamefont
  {Fischer}}, \bibinfo {author} {\bibfnamefont {S.~H.}\ \bibnamefont {Kienle}},
  \ and\ \bibinfo {author} {\bibfnamefont {M.}~\bibnamefont {Freyberger}},\
  }\href {\doibase 10.1103/PhysRevA.61.032306} {\bibfield  {journal} {\bibinfo
  {journal} {Phys. Rev. A}\ }\textbf {\bibinfo {volume} {61}},\ \bibinfo
  {pages} {032306} (\bibinfo {year} {2000})}\BibitemShut {NoStop}%
\bibitem [{\citenamefont {{Hannemann \emph{et
  al}}}(2002)}]{SelfLearningExperimental}%
  \BibitemOpen
  \bibfield  {author} {\bibinfo {author} {\bibfnamefont {T.}~\bibnamefont
  {{Hannemann \emph{et al}}}},\ }\href {\doibase 10.1103/PhysRevA.65.050303}
  {\bibfield  {journal} {\bibinfo  {journal} {Phys. Rev. A}\ }\textbf {\bibinfo
  {volume} {65}},\ \bibinfo {pages} {050303} (\bibinfo {year}
  {2002})}\BibitemShut {NoStop}%
\bibitem [{\citenamefont {Blume-Kohout}()}]{BayesianTomography}%
  \BibitemOpen
  \bibfield  {author} {\bibinfo {author} {\bibfnamefont {R.}~\bibnamefont
  {Blume-Kohout}},\ }\href {\doibase 10.1088/1367-2630/12/4/043034} {\bibfield
  {journal} {\bibinfo  {journal} {New J.\ Phys.}\ }\textbf {\bibinfo {volume}
  {12}},\ \bibinfo {pages} {043034}}\BibitemShut {NoStop}%
\bibitem [{\citenamefont {Doucet}\ \emph {et~al.}(2001)\citenamefont {Doucet},
  \citenamefont {{de Freitas}},\ and\ \citenamefont {Gordon}}]{SMCBook}%
  \BibitemOpen
  \bibinfo {editor} {\bibfnamefont {A.}~\bibnamefont {Doucet}}, \bibinfo
  {editor} {\bibfnamefont {N.}~\bibnamefont {{de Freitas}}}, \ and\ \bibinfo
  {editor} {\bibfnamefont {N.}~\bibnamefont {Gordon}},\ eds.,\ \href@noop {}
  {\emph {\bibinfo {title} {Sequential Monte Carlo in Paractice}}}\ (\bibinfo
  {publisher} {Springer-Verlag},\ \bibinfo {year} {2001})\BibitemShut {NoStop}%
\bibitem [{\citenamefont {Blume-Kohout}\ and\ \citenamefont
  {Hayden}(2006)}]{BayesianOptimality}%
  \BibitemOpen
  \bibfield  {author} {\bibinfo {author} {\bibfnamefont {R.}~\bibnamefont
  {Blume-Kohout}}\ and\ \bibinfo {author} {\bibfnamefont {P.}~\bibnamefont
  {Hayden}},\ }\href@noop {} {\  (\bibinfo {year} {2006})},\ \Eprint
  {http://arxiv.org/abs/quant-ph/0603116} {arXiv:quant-ph/0603116} \BibitemShut
  {NoStop}%
\bibitem [{\citenamefont {{Gross \emph{et al}}}(2010)}]{CompressedSensing}%
  \BibitemOpen
  \bibfield  {author} {\bibinfo {author} {\bibfnamefont {D.}~\bibnamefont
  {{Gross \emph{et al}}}},\ }\href {\doibase 10.1103/PhysRevLett.105.150401}
  {\bibfield  {journal} {\bibinfo  {journal} {Phys.\ Rev.\ Lett.}\ }\textbf
  {\bibinfo {volume} {105}},\ \bibinfo {pages} {150401} (\bibinfo {year}
  {2010})}\BibitemShut {NoStop}%
\bibitem [{\citenamefont {Patra}(2007)}]{ExactInformation}%
  \BibitemOpen
  \bibfield  {author} {\bibinfo {author} {\bibfnamefont {M.~K.}\ \bibnamefont
  {Patra}},\ }\href {http://stacks.iop.org/1751-8121/40/i=35/a=011} {\bibfield
  {journal} {\bibinfo  {journal} {J.\ Phys.\ A}\ }\textbf {\bibinfo {volume}
  {40}},\ \bibinfo {pages} {10887} (\bibinfo {year} {2007})}\BibitemShut
  {NoStop}%
\end{thebibliography}%

\end{document}